\documentclass[nopacs,preprintnumbers,amsmath,amssymb,superscriptaddress]{revtex4}
\usepackage[dvips]{graphicx}
\usepackage{amssymb}
\usepackage{amsmath}
\usepackage{graphicx,color}

\usepackage{times}

\usepackage{enumerate}
\usepackage{mathtools}
\usepackage{stmaryrd}

\newcommand{\moy}[1]{\left\langle #1 \right\rangle}

\newcommand{\ex}[1]{\mathrm{e}^{#1}}

\newcommand{\dd}[0]{\mathrm{d}}
\newcommand{\ii}[0]{\mathrm{i}}
\newcommand{\kk}[0]{\boldsymbol{k}}
\newcommand{\zz}[0]{\mathbf{0}}

\newcommand{\ee}[0]{\boldsymbol{e}}
\newcommand{\rr}[0]{\boldsymbol{r}}

\newcommand{\sss}[0]{\boldsymbol{s}}

\newcommand{\sgn}[0]{\mathrm{sgn}}

\begin{document}

\title{Propagators of random walks on comb lattices of arbitrary dimension}

\author{Pierre Illien}
\address{Rudolf Peierls Centre for Theoretical Physics, University of Oxford, Oxford OX1 3NP, UK}
\address{Department of Chemistry, The Pennsylvania State University, University Park, Pennsylvania 16802, USA}

\author{Olivier B\'enichou}
\address{Laboratoire de Physique Th\'eorique de la Mati\`ere Condens\'ee, UPMC, CNRS UMR 7600, Sorbonne Universit\'es, 4 Place Jussieu, 75252 Paris Cedex 05, France}

\begin{abstract}

We study diffusion on  comb lattices  of arbitrary dimension. Relying on the loopless structure of these lattices and using first-passage properties, we obtain exact and explicit formulae for the Laplace transforms of the propagators associated to nearest-neighbour random walks in both cases where either the first or the last point of the random walk is on the backbone of the lattice, and where the two extremities are arbitrarily chosen. As an application, we compute the mean-square displacement of a random walker on a comb of arbitrary dimension. We also propose an alternative and consistent approach of the problem using a master equation description, and obtain simple and generic expressions of the propagators. This method is more general and is  extended to study the propagators of random walks on more complex comb-like structures. In particular, we study the case of a two-dimensional comb lattice with teeth of finite length.

\end{abstract}

\maketitle

\tableofcontents


\section{Introduction}


Diffusion of particles in systems with  geometrical constraints, such as fractal  or disordered lattices, has motivated a large amount of theoretical work in the past decades \cite{Ben-avraham,{Bouchaud1990}}. This question is central in many physical and biological systems, and arises for instance in the study of transport in porous media, polymer mixtures or living cells. 

Comb-like lattices have received a particular interest, as they minimally reproduce the main features of percolation clusters, or more generally of loopless structures: a line (called thereafter the \emph{backbone}) spans  from one end of the system to the other, to which finite or infinite structures (the \emph{teeth}) are connected. As an example, the simplest two-dimensional comb structure is obtained from a regular two-dimensional square lattice by removing all the lines parallel to the $x$-axis expect from this axis itself. Several extensions of this lattice, such as combs with teeth of random length or generalized higher-dimensional combs, have been studied \cite{Ben-avraham}.

The simplicity of such lattices makes it possible to derive numerous exact results, among which the mean square displacement of an isolated random walker \cite{Weiss1986, {Gan1987},{Pottier1995},{Mendez2015a}}, of a tracer particle in a crowded environment with excluded-volume interactions \cite{Benichou2015a}, first-passage time and survival probability \cite{Kahng1989}, or occupation times statistics in subdomains of the lattice \cite{Rebenshtok2013}. Continuous descriptions of comblike structures have also been proposed, and used to study anomalous diffusion and the influence of drift on the diffusion properties \cite{Arkhincheev1991,Arkhincheev2002, {Iomin2016}}. Finally, in addition to their theoretical interest, comb lattices have been successfully used to model different real systems, among which we can cite spiny dendrites \cite{Mendez2013} or dendronized polymers \cite{Frauenrath2005a}.

We will focus here on the propagators associated to the random walk, namely the probability for a random walker to be at a given site at a given time knowing its starting point. When the starting and arrival points coincide and are located on the backbone, this quantity has been computed for combs of arbitrary dimension \cite{Gerl1986a}. On two-dimensional combs, the propagators of random walks starting from or arriving to the backbone have also been calculated and studied asymptotically \cite{Bertacchi2003,Bertacchi2006a}. In this paper, relying on similar methods, we generalize these results to combs of arbitrary dimension, and obtain the Laplace transforms of the propagators between  arbitrary points of a two-dimensional comb. We also introduce an alternative derivation of these propagators that relies on a master equation formulation of the problem, and that yields a surprisingly simple and explicit formula for the propagator that holds for any starting and arrival points. We finally give a few possible extensions of this method to other lattices  (three-dimensional comb with infinite teeth, two-dimensional comb with finite teeth). The master equation appears to be a powerful and efficient formulation of the problem, allowing one to study random walks on generalized comb-like lattices.\\

The paper is organized as follows: in Section \ref{Sec2}, we present useful notations and fundamental relations that will be used throughout the paper. In Section \ref{prop_pt_bb}, we obtain the propagators of a random walk on a comb of arbitrary dimension when either the first or the last point of the random walk  is on the backbone of the lattice. As a physical application of this computation, we also derive the mean-square displacement of a random walker along the backbone of a comb of arbitrary dimension. In Section \ref{Sec4}, we focus on the two-dimensional comb and derive the propagators for any starting and arrival points. Finally, in Section \ref{ME}, we present a master equation description of the problem, which is consistent with the previous approaches and which allows one to study more complex structures, namely a three-dimensional comb and a two-dimensional comb with finite teeth.

\section{Definitions and basic relations}
\label{Sec2}

 For any time-dependent function $\phi_t$, we define the associated generating function (or discrete Laplace transform) by
\begin{equation}
\label{DLTdef}
\widehat{\phi}(\xi)=\sum_{t=0}^\infty \phi_t \xi^t.
\end{equation}
For any space-dependent function $\psi(\rr)$, we define its Fourier transform
$\widetilde{\psi}(\kk)$ by
\begin{equation}
\label{ }
\widetilde{\psi}(\kk) = \sum_{\rr} \ex{\ii \kk\cdot \rr} \psi(\rr),
\end{equation}
where the sum over $\rr$ runs over all lattice sites. We denote by $P_t(\rr|\rr_0)$ the probability for the random walker to be at site $\rr$ at time $t$ knowing that it was at site $\rr_0$ at time $0$ (this quantity will also be called the \emph{propagator} of the random walk). The associated generating function is then 
\begin{equation}
\label{ }
\widehat{P}(\rr|\rr_0;\xi) = \sum_{t=0}^{\infty}P_t(\rr|\rr_0) \xi^t.
\end{equation}
Following \cite{Woess2000}, we call $T(\rr)$ the minimum of the all the time steps (with $t=0$ included) where the random walker is at site $\rr$. We define the first-passage time density $F_t(\rr|\rr_0)$ as
\begin{equation}
\label{ }
F_t(\rr|\rr_0) = \mathrm{Prob}[T(\rr)=t],
\end{equation}
and deduce the associated generating function $\widehat{F}(\rr|\rr_0;\xi)$ with Eq. (\ref{DLTdef}). From the definition of $T(\rr)$, it is obvious that $\widehat{F}(\rr|\rr;\xi)=1$ for any site $\rr$. When $\rr\neq\rr_0$, it is straightforward to establish the following \emph{renewal equation}, relating the first-passage time density and the propagators:
\begin{equation}
\label{renewal}
\widehat{P}(\rr|\rr_0;\xi)  = \widehat{P}(\rr|\rr;\xi) \widehat{F}(\rr|\rr_0;\xi).
\end{equation}
In the particular case where $\rr=\rr_0$, the propagator $\widehat{P}(\rr|\rr_0;\xi)$ may also be related to the first-passage time densities through \cite{Woess2000}:
\begin{equation}
\label{renewalcoincidant}
\widehat{P}(\rr|\rr;\xi) = \frac{1}{1-\xi\sum_{\rr'} \widehat{F}(\rr|\rr';\xi) p(\rr'|\rr)},
\end{equation}
where $p(\rr'|\rr)$ is the probability to jump from $\rr$ to $\rr'$ in a single time step.

We finally notice that comb lattices are examples of tree-like structures, which means that for two arbitrary nodes $\boldsymbol{r}$ and $\boldsymbol{r}'$ separated by a distance $d(\boldsymbol{r},\boldsymbol{r}')$, there exists only one path of length $d(\boldsymbol{r},\boldsymbol{r}')$, denoted by $\gamma(\boldsymbol{r},\boldsymbol{r}')$. This property implies the following relation \cite{Woess2000}:
\begin{equation}
\label{tree}
\widehat{F}(\boldsymbol{r}|\boldsymbol{r}';\xi) = \widehat{F}(\boldsymbol{r}|\boldsymbol{r}'';\xi)\widehat{F}(\boldsymbol{r}''|\boldsymbol{r}';\xi),
\end{equation}
which holds for any site $\boldsymbol{r}''$ belonging to the path $\gamma(\boldsymbol{r},\boldsymbol{r}')$. In other words, on a loopless lattice, the Laplace transform of the first-passage time density of a random walk between two given points can be decomposed by considering intermediate points belonging to the shortest path between the starting and arrival points.

\section{Random walk on a $d$-comb}
\label{prop_pt_bb}

\subsection{Definition of the lattice}

\begin{figure}
\begin{center}
\includegraphics[width=6cm]{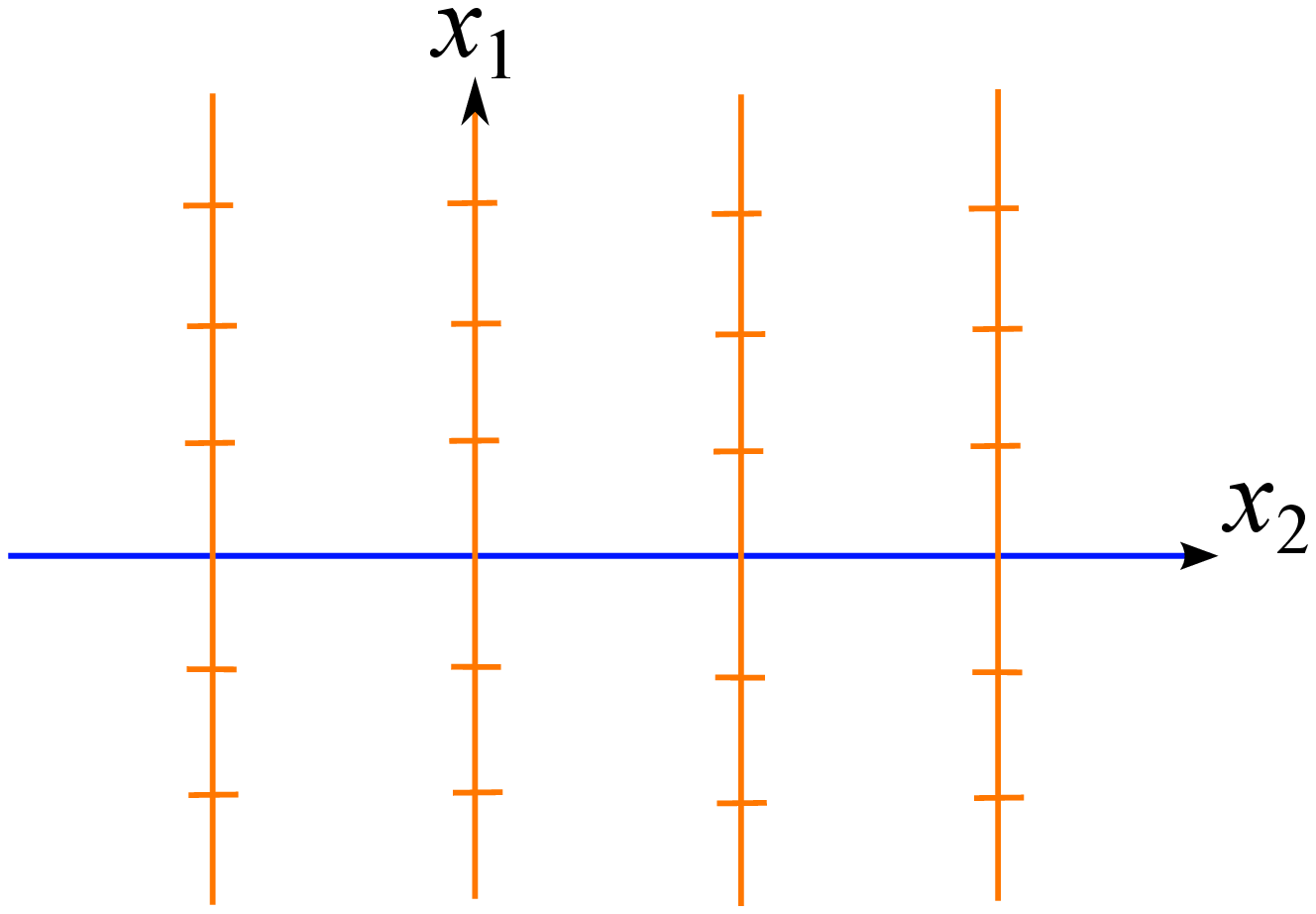}\\
\includegraphics[width=8cm]{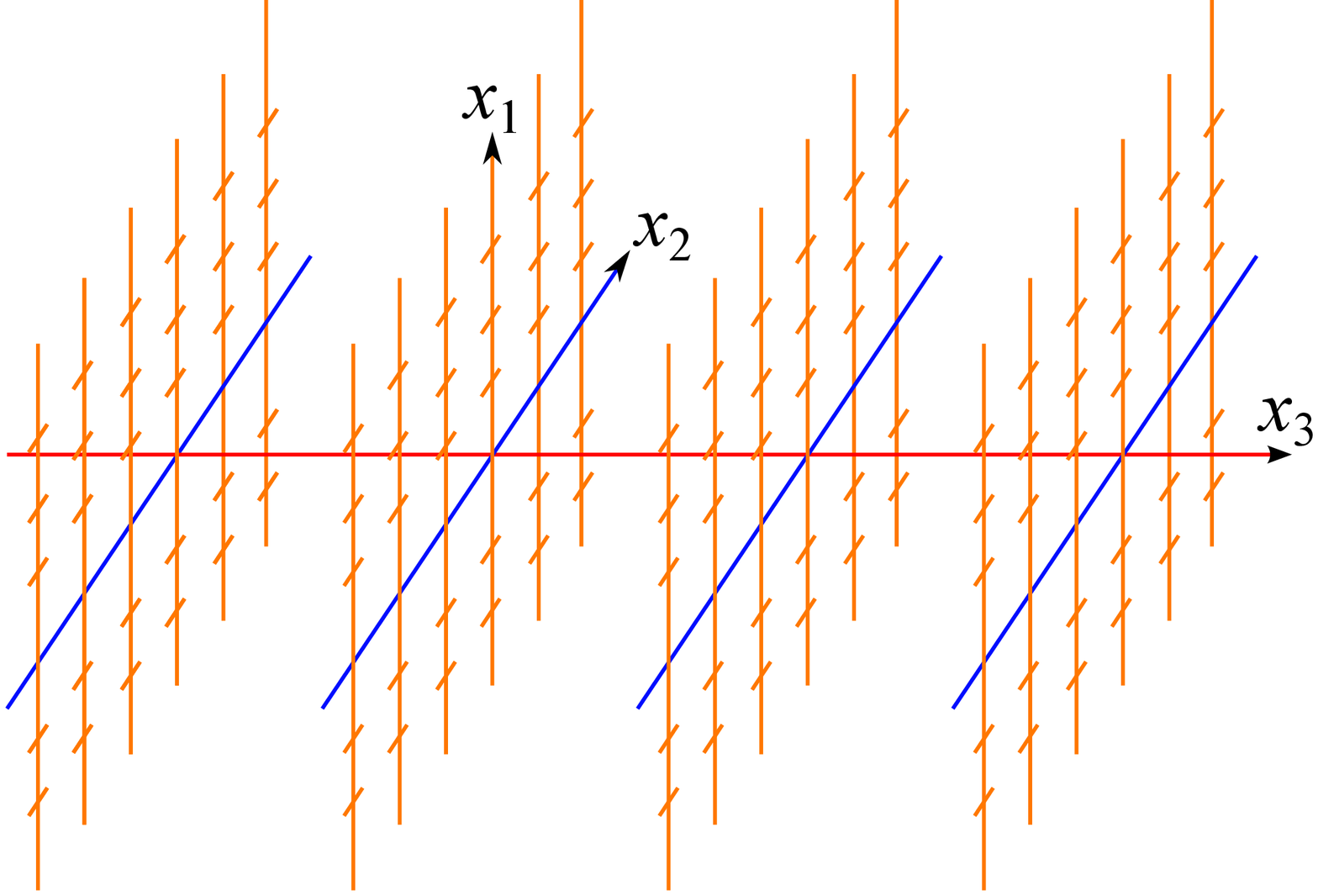}
\caption{The structures of the two-dimensional (top) and three-dimensional (bottom) combs.}
\label{fig:combs}
\end{center}
\end{figure}

For $d\geq 1$, the $d$-dimensional comb (or $d$-comb), denoted by $\mathbf{C}_d$,  is defined recursively as follows: $\mathbf{C}_d$ is obtained from $\mathbf{C}_{d-1}$ by attaching to each site of $\mathbf{C}_{d-1}$ an infinite line of integers, $\mathbf{C}_1$ being the one-dimensional regular lattice. Equivalently, $\mathbf{C}_d$ can be built starting from a one-dimensional regular lattice whose sites are attached to the backbone of a copy of $\mathbf{C}_{d-1}$. We represent the $2$-comb and the $3$-comb  on Fig. \ref{fig:combs}.

The $d$ unit vectors pointing out from a backbone site of $\mathbf{C}_d$ actually coincide with the backbones of lower-dimensional combs. We choose the following convention: the unit vector aligned with the backbone of a copy of $\mathbf{C}_j$ will be denoted by $\ee_j$. Therefore, the direction of the backbone of $\mathbf{C}_d$ coincides with that of the unit vector $\ee_d$.

\subsection{Propagator of a random walk with ending point on the backbone}
\label{endingpointbb}

In this section, we compute the generating function $\widehat{P}(\zz|\rr;\xi)$ associated to a random walk with an arbitrary starting point $\rr=\sum_{i=1}^d r_i \ee_i$ and arriving at the origin of the lattice (i.e. on the backbone). Using the renewal equation (Eq. (\ref{renewal})), we write
\begin{equation}
\label{ }
\widehat{P}(\zz|\rr;\xi) = \widehat{P}(\zz|\zz;\xi) \widehat{F}(\zz|\rr;\xi)
\end{equation}
In what follows, we study separately the two generating functions $ \widehat{P}(\zz|\zz;\xi)$ and $\widehat{F}(\zz|\rr;\xi)$. The generating function  $G_d(\xi) \equiv \widehat{P}(\zz|\zz;\xi)$ (propagator with coinciding starting and arrival points on the backbone) has already been studied and is defined recursively \cite{Gerl1986a}:
\begin{equation}
\label{Gd_recurs}
G_d(\xi)=\frac{d}{\sqrt{\left( 1+\frac{d-1}{G_{d-1}(\xi)}\right)^2-\xi^2}},
\end{equation}
with $G_1(\xi)=1/\sqrt{1-\xi^2}$ \cite{Hughes1995}. The following expression of $G_2(\xi)$ will be used several times throughout this paper:
\begin{equation}
\label{G2}
G_2(\xi) = \frac{\sqrt{2}}{\sqrt{1-\xi^2+\sqrt{1-\xi^2}}}.
\end{equation}

The generating function associated to the first-passage time density $\widehat{F}(\zz|\rr;\xi)$ can be calculated using Eq. (\ref{tree}). Decomposing the path from $\rr$ to $\zz$ as follows
\begin{equation}
\label{ }
\rr  \longrightarrow \rr-r_1 \ee_1 \longrightarrow \rr-r_1\ee_1 - r_2\ee_2 \longrightarrow \dots \longrightarrow r_d\ee_d \longrightarrow \zz,
\end{equation}
we obtain
\begin{equation}
\label{ }
\widehat{F}(\zz|\rr;\xi)=\widehat{F}(\zz|r_d \ee_d;\xi) \dots \widehat{F}(\rr - r_1\ee_1-r_2\ee_2|\rr -r_1\ee_1;\xi)  \widehat{F}(\rr - r_1\ee_1 | \rr ;\xi).
\end{equation}
Noticing that the $j$-th step of the decomposed path takes place on the backbone of a $j$-comb,  defining $f_d(\xi)$ as
\begin{equation}
\label{ }
f_d(\xi) \equiv \widehat{F}(\zz|\ee_d;\xi),
\end{equation}
and using again Eq. (\ref{tree}), we get
\begin{eqnarray}
\widehat{F}(\zz|\rr;\xi) & = & f_d(\xi)^{|r_d|}\dots f_2(\xi)^{|r_2|}  f_1(\xi)^{|r_1|}\\
 & = & \prod_{j=1}^d f_j(\xi)^{|r_j|}.
\end{eqnarray}

In what follows, we establish a recurrent definition of $f_d(\xi)$. Starting from its definition and partitioning over the first step of the walk, one writes
\begin{eqnarray}
f_d(\xi) & = & \widehat{F}(\zz|\ee_d;\xi) \\
 & = & \xi \sum_{\boldsymbol{w}}  \widehat{F}(\zz|\boldsymbol{w};\xi) p(\boldsymbol{w}|\ee_d),
\end{eqnarray}
where we denote by $p(\rr|\rr_0)$ the probability for the random walker to jump from site $\rr_0$ to site $\rr$ in a single step. From site $\ee_d$, the random walker has $2d$ neighboring sites on which it may jump equiprobably, so that we get
\begin{eqnarray}
f_d(\xi) & = & \xi \left[  \widehat{F}(\zz|\zz;\xi) p(\zz|\ee_d)+ \widehat{F}(\zz|2 \ee_d;\xi) p(2\ee_d|\ee_d)  + 2 \sum_{j=1}^{d-1}  \widehat{F}(\zz|\ee_d+\ee_j;\xi) p(\ee_d+\ee_j|\ee_d)   \right]\\
 & = & \frac{\xi}{2d}  \left[  1+ f_d(\xi)^2 + 2 \sum_{j=1}^{d-1}  \widehat{F}(\zz|\ee_d+\ee_j;\xi) \right], \label{fd1}
\end{eqnarray}
where we used $ \widehat{F}(\zz|\zz;\xi) =1$ and Eq. (\ref{tree}) to write $\widehat{F}(\zz|2 \ee_d;\xi)= f_d(\xi)^2$. The generating function $ \widehat{F}(\zz|\ee_d+\ee_j;\xi) $ can be expressed in terms of the $f_j(\xi)$ functions by using Eq. (\ref{tree}):
\begin{eqnarray}
\label{ }
 \widehat{F}(\zz|\ee_d+\ee_j;\xi) &=& \widehat{F}(\zz|\ee_d;\xi)   \widehat{F}(\ee_d|\ee_d+\ee_j;\xi) \\
 &=& f_d(\xi) f_j(\xi).
\end{eqnarray}
From Eq. (\ref{fd1}), we thus obtain that $f_d(\xi)$ is the solution of the following second-order equation:
\begin{equation}
\label{ }
f_d(\xi)^2+2 f_d(\xi) \left(  \sum_{j=1}^{d-1} f_j(\xi) - \frac{d}{\xi} \right) +1=0.
\end{equation}
Selecting the solution fulfilling the condition $f_d(0)=0$, we finally obtain the following expression:
\begin{equation}
\label{f_recurs}
f_d(\xi) = \frac{d}{\xi} - \sum_{j=1}^{d-1} f_j(\xi) - \sqrt{\left( \frac{d}{\xi} - \sum_{j=1}^{d-1} f_j(\xi)\right)^2-1}.
\end{equation}
Consequently, the generating function $f_d(\xi)$ can be computed recursively, starting with the known expression of $\widehat{F}(\zz|\ee_1;\xi)$ on a one-dimensional lattice  \cite{Hughes1995}:
\begin{equation}
\label{f1}
 f_1(\xi)=\frac{1-\sqrt{1-\xi^2}}{\xi}. 
\end{equation}
 In particular, for $d=2$, one retrieves the result previously obtained in \cite{Bertacchi2003}:
 \begin{equation}
\label{f2}
f_2(\xi) = \frac{1}{\xi} \left( 1+\sqrt{1-\xi^2}  -\sqrt{2} \sqrt{1-\xi^2+\sqrt{1-\xi^2}}\right).
\end{equation}

Although there is no explicit expression of the functions $f_d(\xi)$ that can be deduced from the recursive definition in Eq. (\ref{f_recurs}), one can show that the limit of  $\xi\to1$ (i.e.  long-time limit),
\begin{equation}
\label{FPTDdev}
f_d(\xi) \underset{\xi\to1}{=} 1-2^{1-1/2^d}(1-\xi)^{1/2^d} + \mathcal{O}\left[(1-\xi)^{1/2^{d-1}}\right].
\end{equation}
Using a Tauberian theorem \cite{Hughes1995}, we then find the long-time expansion of the first-passage time densities (FPTD)
\begin{equation}
\label{ }
F_t(\zz|\ee_d) \underset{t\to\infty}{\sim} \frac{2^{1-1/2^d}}{\Gamma(1-1/2^d)} \frac{1}{t^{1+1/2^d}}.
\end{equation}
In particular, for $d=1$, we retrieve the well-known power-law decrease of the FPTD for a one-dimensional simple symmetric random walk \cite{Hughes1995} : $F_t(\zz|\ee_1) \propto_{t\to\infty} 1/t^{3/2}$.

Finally, the generating function $\widehat{P}(\zz|\rr;\xi)$ is given by the following relation:
\begin{equation}
\label{P_arrive_bb}
\widehat{P}(\zz|\rr;\xi)=G_d(\xi) \prod_{j=1}^{d} f_j(\xi)^{|r_j|},
\end{equation}
where $f_j(\xi)$ is defined recursively by Eqs. (\ref{f_recurs}) and (\ref{f1}), and where  $G_d(\xi)$ is defined recursively by Eq. (\ref{Gd_recurs}) together with the one-dimensional expression $G_1(\xi)=1/\sqrt{1-\xi^2}$.

\subsection{Propagator of a random walk starting from the backbone}

The  propagator $\widehat{P}(\rr|\zz;\xi)$ associated to a random walk starting from the backbone can be deduced straightforwardly from the previous calculation using the following relation, that will be referred to as the \emph{reversibility} property \cite{Bertacchi1999}:
\begin{equation}
\label{rev}
\frac{1}{\nu(\rr)} \widehat{P}(\rr|\rr';\xi)= \frac{1}{\nu(\rr')} \widehat{P}(\rr'|\rr;\xi),
\end{equation}
where $\nu(\rr)$ is the number of neighbors (or the degree) of site $\rr$. More precisely, for a $d$-comb, it is given by
\begin{equation}
\label{ }
\nu(\rr)=\begin{cases}
  2d    & \text{if $\rr=\zz$}, \\
   2\,  \mathrm{min} \{ j\in [1,d] | r_j\neq 0   \}   & \text{otherwise}.
\end{cases}
\end{equation}
Finally, we find
\begin{equation}
\label{reverse}
\widehat{P}(\rr|\zz;\xi)  =  \frac{\nu(\rr)}{\nu(\zz)}  \widehat{P}(\zz|\rr;\xi) \\
 =  \frac{\nu(\rr)}{2d} G_d(\xi) \prod_{j=1}^{d} f_j(\xi)^{|r_j|},
\end{equation}
where we used the result from Eq. (\ref{P_arrive_bb}).

\subsection{Mean-square displacement of a random walker}

As an application of the result presented in the previous section, we aim to compute the mean-square displacement (MSD) of a random walker along the backbone of  a $d$-comb. Its Laplace transform  is related to the Fourier-Laplace transform of the propagator (or moment generating function) through the relation
\begin{equation}
\label{ }
\widehat{\moy{x^2}}(\xi) =  - \left[ \frac{\partial^2}{\partial {k_d}^2} \widetilde{\widehat{P}}(\kk|\zz;\xi) \right]_{\kk=\zz}.
\end{equation}
We  first compute the Fourier transform of the propagator given by Eq. \eqref{reverse}:
\begin{equation}
\label{ }
\widetilde{\widehat{P}}(\kk|\zz;\xi) = \frac{G_d(\xi)}{2d} \sum_{r_1,\dots,r_d}\ex{\ii \kk \cdot \rr} \nu(\rr) \prod_{j=1}^d f_j(\xi)^{|r_j|}.
\end{equation}
The sum in the above equation runs over the lattice sites. We split it depending on the values of the connectivity $\nu$. We denote by $\mathcal{D}_{2j}$ the ensemble of the lattice sites whose connectivity is $2j$:
\begin{equation}
\label{ }
\mathcal{D}_{2j} = 
\begin{cases}
    \{   \rr |   |r_1|\geq1, r_2,\dots,r_d \in \mathbb{Z}    \}     & \text{if $j=1$}, \\
        \{   \rr |   r_1=\dots=r_{d-1}=0, r_d \in \mathbb{Z}    \}     & \text{if $j=d$}, \\
    \{   \rr |   r_1=\dots=r_{j-1}=0, |r_j| \geq 1, r_3,\dots,r_d \in \mathbb{Z}    \}     & \text{otherwise}.
\end{cases}
\end{equation}
The Fourier-Laplace transform of the propagators then becomes
\begin{equation}
\label{PDj}
\widetilde{\widehat{P}}(\kk|\zz;\xi) = \frac{G_d(\xi)}{2d} \left\{  2 \sum_{\rr\in \mathcal{D}_2}\ex{\ii \kk \cdot \rr}\prod_{j=1}^d f_j(\xi)^{|r_j|}+ \dots+ 2d \sum_{\rr\in \mathcal{D}_{2d}}\ex{\ii \kk \cdot \rr} \prod_{j=1}^d f_j(\xi)^{|r_j|} \right\} .
\end{equation}
Using the relation
\begin{equation}
\label{ }
\sum_{r_j}  \ex{\ii k_j r_j}f_j^{|r_j|} = \frac{1-{f_j}^2}{1+{f_j}^2-2f_j\cos k_j},
\end{equation}
it is straightforward to compute the sums  in the rhs of Eq. \eqref{PDj}. Taking $k_1=\dots=k_{d-1}=0$ and derivating with respect to $k_d$, one obtains
\begin{equation}
\label{ }
\widehat{\moy{x^2}}(\xi) =  \frac{G_d}{d} \frac{f_d(1+f_d)}{(1-f_d)^3}\left\{ 2 \frac{2f_1}{1-f_1} \prod_{j=2}^{d-1} \frac{1+f_j}{1-f_j} + 4 \frac{2f_2}{1-f_2} \prod_{j=3}^{d-1} \frac{1+f_j}{1-f_j} +\dots+2d \right\}.
\end{equation}
Finally, using the $\xi \to 1$ expansion of the FPTD $f_j$ (Eq. \eqref{FPTDdev}), and deducing from the recursive definition of $G_d$ (Eq. \eqref{Gd_recurs}) that $G_d(\xi) \sim_{\xi\to1} 2^{1/2^d-1} d/(1-\xi)^{1/2^d}$, we finally obtain the following expansion of $\widehat{\moy{x^2}}(\xi)$ in the limit of $\xi\to1$:
\begin{equation}
\label{ }
\widehat{\moy{x^2}}(\xi) \underset{\xi \to 1}{\sim} \frac{2^{\frac{1}{2^{d-1}}-1}}{(1-\xi)^{1+\frac{1}{2^{d-1}}}}.
\end{equation}
Using a Tauberian theorem \cite{Hughes1995}, we get the long-time limit limit of the MSD of a random walker along the backbone of a $d$-comb:
\begin{equation}
\label{MSDarbitraryd}
\moy{x^2}(t) \underset{t \to \infty}{\sim} \frac{2^{\frac{1}{2^{d-1}}-1}}{\Gamma\left( 1+\frac{1}{2^{d-1}}  \right)} t ^{\frac{1}{2^{d-1}}}.
\end{equation}
For a one-dimensional lattice, we retrieve $\moy{x^2}_{\text{1D}} \sim t$ \cite{Hughes1995}. For $d=2$, we retrieve the result proved by Weiss and Havlin \cite{Weiss1986}: $\moy{x^2}_{\text{2D}} \sim \sqrt{2t/\pi}$. The result given in Eq. \eqref{MSDarbitraryd}  indicates that diffusion along the backbone of a comb is anomalous for $d\geq 2$, and that the mean-square displacement along the backbone grows slower and slower when the dimension of the comb increases. This can be understood as a consequence of the increasing time lost by the random walker on the structures branched to the backbone, whose dimension and complexity increase when $d$ increases.

\section{Random walk between two arbitrary points of $\mathbf{C}_2$}
\label{Sec4}

The general situation where the initial or final point of the random walk does not belong to the backbone of the comb requires more attention. In this section, we will give explicit expressions of the generating functions associated to the generic propagators of a random walk on $\mathbf{C}_2$, and we will consider separately two cases: (i) the situation where the initial and final points of the random walk (respectively $\sss$ and $\rr$) do not belong to the same half-tooth (i.e. $r_2 \neq s_2$ or $r_2=s_2$ and $\sgn(r_1)\neq\sgn(s_1)$), which can be deduced straightforwardly from the results presented in the previous section; (ii) the situation where the initial and starting points belong to the same half-tooth. 

\subsection{First case: $\rr$ and $\sss$ do not belong to the same half-tooth}

In this section, we deduce from the previous calculations the expression of the propagator $\widehat{P}(\rr|\sss;\xi)$ where $\rr$ and $\sss$ do not belong to the same half-tooth. More precisely, this means that the shortest path from $\sss$ to $\rr$ includes at least one point of the backbone (or, equivalently, that $r_2 \neq s_2$ or $r_2=s_2$ and $\sgn(r_1)\neq\sgn(s_1)$). Using the renewal equation, we write
\begin{equation}
\label{ }
\widehat{P}(\rr|\sss;\xi) = \widehat{F}(\rr|\sss;\xi) \widehat{P}(\rr|\rr;\xi).
\end{equation}
We decompose the path from $\sss$ to $\rr$ as follows:
\begin{equation}
\label{path_decomp}
\sss \longrightarrow s_2 \ee_2 \longrightarrow r_2\ee_2 \longrightarrow \rr,
\end{equation}
and use Eq. (\ref{tree}) to write
\begin{eqnarray}
\widehat{P}(\rr|\sss;\xi) & = & \widehat{F}(\rr|r_2\ee_2;\xi) \widehat{F}(r_2 \ee_2 |s_2\ee_2;\xi) \widehat{F}(s_2 \ee_2 |\sss;\xi) \widehat{P}(\rr |\rr;\xi) \\
 & = &  \widehat{F}(\rr|r_2\ee_2;\xi)f_2(\xi)^{|r_2-s_2|}  f_1(\xi)^{|s_1|}  \widehat{P}(\rr |\rr;\xi).
 \label{noname}
\end{eqnarray} 
The renewal equation between points $r_2\ee_2$ and $\rr$ yields
\begin{equation}
\label{ }
\widehat{F}(\rr|r_2\ee_2;\xi)\widehat{P}(\rr |\rr;\xi) = \widehat{P}(\rr |r_2 \ee_2;\xi),
\end{equation}
and, using Eq. (\ref{reverse}),
\begin{eqnarray}
\label{ }
\widehat{F}(\rr|r_2\ee_2;\xi)\widehat{P}(\rr |\rr;\xi) &=& \frac{\nu(\rr)}{4} \widehat{P}(r_2\ee_2 |\rr;\xi)\\
 &=& \frac{\nu(\rr)}{4} G_2(\xi)f_1(\xi)^{|r_1|}. \label{noname2}
 \end{eqnarray}
Finally, combining Eqs. (\ref{noname}) and (\ref{noname2}), we find the following expression for the generic propagator $\widehat{P}(\rr|\sss;\xi)$:
\begin{equation}
\label{arbitraryC2}
\widehat{P}(\rr|\sss;\xi)=\frac{\nu(\rr)}{4} G_2(\xi) f_2(\xi)^{|r_2-s_2|}  f_1(\xi)^{|r_1|+|s_1|}.
\end{equation}


This result can be easily generalized to a $d$-dimensional comb, in the particular case where the shortest path between $\sss$ and $\rr$ contains at least one point from the backbone (i.e. $r_d\neq s_d$ or $r_d = s_d$ and $\sgn(r_{d-1}) \neq \sgn(r_{d-1})$). The propagator of the random walk from $\sss$ to $\rr$ is then given by
\begin{equation}
\label{ }
\widehat{P}(\rr|\sss;\xi)=\frac{\nu(\rr)}{2d} G_d(\xi) f_d(\xi)^{|r_d-s_d|} \prod_{j=1}^{d-1} f_j(\xi)^{|r_j|+|s_j|}.
\end{equation}

\subsection{Second case: $\rr$ and $\sss$ belong to the same half-tooth}

In the situation where $\rr$ and $\sss$ are two distinct points belonging to the same half-tooth, the path from $\sss$ to $\rr$ cannot be decomposed as in Eq. (\ref{path_decomp}) as it does not contain any point of the backbone, and the results presented in Section \ref{prop_pt_bb} cannot be used anymore. With no loss of generality (using Eq. (\ref{rev})), we assume that $|s_1| > |r_1|$ and $s_1,r_1>0$. The renewal equation (Eq. (\ref{renewal})) yields
\begin{eqnarray}
\label{ }
\widehat{P}(\rr|\sss;\xi) &=& \widehat{P}(\rr|\rr;\xi) \widehat{F}(\rr|\sss;\xi)\\
&=& \widehat{P}(\rr|\rr;\xi) f_1(\xi)^{|s_1-r_1|} 
\end{eqnarray}
where we used Eq. (\ref{tree}) to obtain the last equality.  In what follows we compute the propagator $\widehat{P}(\rr|\rr;\xi)$ associated to a random walk starting and arriving at the same point on a tooth of the comb. For $n\geq1$, we define
\begin{equation}
\label{ }
U_n(\xi) = \widehat{P}(n\ee_1|n\ee_1;\xi).
\end{equation}
We relate this propagator to the first-passage time density using Eq. (\ref{renewalcoincidant}):
\begin{equation}
\label{Un}
U_n(\xi)=\frac{1}{1-\xi \sum_{\boldsymbol{w}}    \widehat{F}(n\ee_1|\boldsymbol{w};\xi) p(\boldsymbol{w}|n\ee_1)}.
\end{equation}
The sum in the denominator can be written explicitly:
\begin{eqnarray}
\label{ }
\xi \sum_{\boldsymbol{w}}    \widehat{F}(n\ee_1|\boldsymbol{w};\xi) p(\boldsymbol{w}|n\ee_1)&=& \xi \left[   \frac{1}{2} \widehat{F}(n\ee_1|(n+1)\ee_1;\xi) +   \frac{1}{2} \widehat{F}(n\ee_1|(n-1)\ee_1;\xi) \right] \\
&=&\frac{\xi}{2} \left[  f_1(\xi) +   \widehat{F}(n\ee_1|(n-1)\ee_1;\xi) \right] ,
\end{eqnarray}
and we deduce the following expression for $U_n(\xi)$:
\begin{equation}
\label{Uh}
U_n(\xi)=\frac{1}{1-\frac{\xi}{2}\left[ f_1(\xi) + h_n(\xi)  \right]},
\end{equation}
where we define $h_n(\xi)\equiv \widehat{F}(n\ee_1|(n-1)\ee_1;\xi)$. The previous relation holds for $n\geq1$. To obtain a recurrence relation satisfied by $U_n(\xi)$, we use the reversibility property (Eq. (\ref{rev})), which holds for $n\neq1$:
\begin{equation}
\label{revPn}
\widehat{P}(n \ee_1 |(n-1)\ee_1;\xi) = \widehat{P}((n-1) \ee_1 |n\ee_1;\xi).
\end{equation}
Both sides of Eq. (\ref{revPn}) are calculated using the renewal equation (Eq. (\ref{renewal})):
\begin{equation}
\label{ }
\widehat{P}(n \ee_1 |n\ee_1;\xi) \widehat{F}(n \ee_1 |(n-1)\ee_1;\xi) = \widehat{P}((n-1) \ee_1 |(n-1)\ee_1;\xi)\widehat{F}((n-1) \ee_1 | n \ee_1;\xi),
\end{equation}
and using the definition of $U_n(\xi)$ and $h_n(\xi)$, one gets
\begin{equation}
\label{revUhf}
 U_n(\xi) h_n(\xi) =U_{n-1}(\xi) f_1(\xi) .
\end{equation}
Using Eq. (\ref{Uh}), $h_n(\xi)$ can be written as a function of $U_n(\xi)$:
\begin{equation}
\label{ }
h_n(\xi) = \frac{2}{\xi}\left[ 1-\frac{1}{U_n(\xi)}  \right]-f_1(\xi),
\end{equation}
or, equivalently, using the expression of $f_1(\xi)$  (Eq. (\ref{f1})) and the relation $2/\xi-f_1(\xi)=1/f_1(\xi)$, we get
\begin{equation}
\label{hnUn}
h_n(\xi) = \frac{1}{f_1(\xi)}-\frac{2}{\xi U_n(\xi)}.
\end{equation}
Using this relation in Eq. (\ref{revUhf}), we obtain the following recurrence relation satisfied by $U_n(\xi)$ which holds for $n>1$:
\begin{equation}
\label{recrel}
U_n(\xi) = f_1(\xi)^2 U_{n-1}(\xi) + \frac{2 f_1(\xi)}{\xi}.
\end{equation}
To determine $U_1(\xi)$, we use again the reversibility property (Eq. (\ref{rev})) to write the following equations:
\begin{eqnarray}
\label{ }
\widehat{P}(\ee_1|\zz;\xi) &=& \frac{1}{2} \widehat{P}(\zz|\ee_1;\xi) \\
\widehat{P}(\ee_1|\ee_1;\xi) \widehat{F}(\ee_1|\zz ;\xi)  &=& \frac{1}{2} \widehat{P}(\zz|\zz;\xi) \widehat{F}(\zz|\ee_1 ;\xi) \\
U_1(\xi) h_1(\xi) & =& \frac{1}{2} G_2(\xi)  f_1(\xi) \label{U1rel}.
\end{eqnarray}
Noticing that Eq. (\ref{hnUn}) still holds for $n=1$, we can eliminate $h_1(\xi)$ from Eq. (\ref{U1rel}) and obtain
\begin{equation}
\label{U1}
U_1(\xi) = \frac{1}{2} G_2(\xi) f_1(\xi)^2 + \frac{2}{\xi} f_1(\xi).
\end{equation}
Finally, solving the recurrence relation (Eq. (\ref{recrel})) with the boundary condition from Eq. (\ref{U1}), we get the following explicit expression of $U_n(\xi) = \widehat{P}(n\ee_1 |n\ee_1;\xi)$:
\begin{equation}
\label{ }
U_n(\xi) = \frac{1}{2 } G_2(\xi)  f_1(\xi)^{2n} + \frac{2}{\xi} f_1(\xi)^{2n-1} +G_1(\xi) \left[1-f_1(\xi)^{2(n-1)}\right].
\end{equation}
This yields the following expression for $\widehat{P}(\rr|\sss;\xi)$ when $\rr$ and $\sss$ belong to the same half-tooth:
\begin{equation}
\label{arbitraryC22}
\widehat{P}(\rr|\sss;\xi) = f_1(\xi)^{|s_1-r_1|}  \left\{ \frac{1}{2 } G_2(\xi)  f_1(\xi)^{2r_1} + \frac{2}{\xi} f_1(\xi)^{2r_1-1} +G_1(\xi) \left[1-f_1(\xi)^{2(r_1-1)}\right]  \right\}.
\end{equation}

To summarise, relying on the loopless structure of the lattice and using first-passage properties, we calculated in Section \ref{prop_pt_bb} the propagators and first-passage time densities of a random walk whose starting or ending point is on the backbone of the lattice of a comb of \emph{arbitrary} dimension. In Section \ref{Sec4}, we computed the propagators of random walks with arbitrary starting and ending points on a two-dimensional lattice.

The generalisation of  this calculation to the case of a three-dimensional comb (or even a comb of arbitrary dimension) is too complicated to be presented here. We propose in the next section an alternative and more straightforward calculation of the propagators of a random walk on a comb, which relies on a master-equation formulation of the problem.

\section{A master equation derivation of the propagators}
\label{ME}

In this section, we write the master equation describing the evolution of the propagators $P_t(\rr | \sss)$ on a two-dimensional comb. As the comb lattice is not translation invariant, the master equation will depend on the location of the arrival point $\rr$. Relying on the observation that the comb is an homogeneous lattice with an infinity of particular points arranged along a single line, we show that the problem is completely described by a set of three master equations, from which we compute the Fourier transform of the  propagator $P_t(\rr | \sss)$  \cite{Nieuwenhuizen2004}. We are able to invert these Fourier transforms to retrieve the results obtained in the previous sections, and show that they are all contained in a single and simple formula.

In a second time, we extend this method to the case of a three-dimensional comb with infinite teeth and to the case of a two-dimensional comb with teeth of finite length, and calculate the propagators of a random walk between two arbitrary points of this structure.

\subsection{Two-dimensional comb}

On a two-dimensional comb, depending on the site $\rr$ occupied by the walker, there are two possible sets of probability jumps:
\begin{enumerate}
\item if $r_1=0$, the walker has a probability $1/4$ to jump into each of the directions $\pm \ee_1$, $\pm \ee_2$.
\item if $r_1\neq 0$, the walker has a probability $1/2$ to jump into each of the directions $\pm \ee_1$.
\end{enumerate}
In order to simplify the notations, in what follows, we drop the explicit $\sss$-dependence of the propagators and will simply consider $P_t(\rr)$ (probability for the walker to be at site $\rr$ at time $t$) with the initial condition:
\begin{equation}
\label{ICP}
P_0(\rr)= \delta_{\rr,\sss}.
\end{equation}

 The master equations of the problem are the following:
 \begin{itemize}
  \item for $|r_1|>1$:
  \begin{equation}
\label{MEbulk}
P_{t+1}(r_1,r_2)=\frac{1}{2}[P_t(r_1+1,r_2)+P_t(r_1-1,r_2)] 
\end{equation}
  \item for $|r_1|=1$:
  \begin{equation}
\label{MEBC1}
P_{t+1}(\pm1,r_2) =\frac{1}{2}P_t(\pm 2, r_2) +\frac{1}{4}P_t(0,r_2)
\end{equation}
  \item for $r_1=0$:
  \begin{equation}
\label{MEBC2}
 P_{t+1}(0,r_2)=\frac{1}{2}[P_t(1,r_2)+P_t(-1,r_2)]  +\frac{1}{4}[P_t(0,r_2+1)+P_t(0,r_2-1)]  
\end{equation}
\end{itemize}

We introduce the following generating functions and Laplace transforms:
\begin{eqnarray}
\mathcal{P}_{\mathrm{b}}(k_2;\xi) & \equiv & \sum_{t=0}^\infty  \sum_{r_2=-\infty}^\infty \ex{\ii k_2 r_2} P_t(0,r_2)\xi^t, \label{def_Pb}\\
\mathcal{P}(k_1,k_2;\xi) & \equiv &  \sum_{t=0}^\infty  \sum_{r_1,r_2=-\infty}^\infty \ex{\ii (k_1 r_1+k_2 r_2)} P_t(r_1,r_2)\xi^t .
\end{eqnarray}
Multiplying Eq. (\ref{MEbulk}) by $\xi^t \ex{\ii (k_1 r_1+k_2 r_2)} $, summing for $t\geq 0$ and on every lattice sites, and using the initial condition (Eq. (\ref{ICP})), one can show  that $\mathcal{P}$ and $\mathcal{P}_{\mathrm{b}}$ are related by
\begin{equation}
\label{rel_P_Pb}
\mathcal{P}(k_1,k_2;\xi)=\frac{\ex{\ii k_1 s_1}\ex{\ii k_2 s_2}+\frac{\xi}{2}(\cos k_2 -\cos k_1)\mathcal{P}_{\mathrm{b}}(k_2;\xi)}{1-\xi \cos k_1}.
\end{equation}
In order to get an equation satisfied by $\mathcal{P}_{\mathrm{b}}(k_2;\xi)$, we integrate each side of this equation over $k_1$ and use the simple relation between $\mathcal{P}$ and $\mathcal{P}_{\mathrm{b}}$:
\begin{equation}
\mathcal{P}_{\mathrm{b}}(k_2;\xi)=\int_0^{2\pi} \frac{\dd k_1}{2\pi} \mathcal{P}(k_1,k_2;\xi).
\end{equation}
This yields the following equation satisfied by $\mathcal{P}_{\mathrm{b}}(k_2;\xi)$:
\begin{eqnarray}
\mathcal{P}_{\mathrm{b}}(k_2;\xi) & = & \ex{\ii k_2 s_2} \int_0^{2\pi} \frac{\dd k_1}{2\pi} \frac{\ex{\ii k_1 s_1}}{1-\xi \cos k_1}   + \frac{\xi}{2} \mathcal{P}_{\mathrm{b}}(k_2;\xi)  \int_0^{2\pi}  \frac{\dd k_1}{2\pi} \frac{\cos k_2-\cos k_1}{1-\xi \cos k_1}     \\
& = & \ex{\ii k_2 s_2} I(s_1)   + \frac{\xi}{2} \mathcal{P}_{\mathrm{b}}(k_2;\xi) \cos (k_2) \, I(0)  - \frac{\xi}{2} \mathcal{P}_{\mathrm{b}}(k_2;\xi)I(1) \label{PbI},
\end{eqnarray}
where we define the integral
\begin{equation}
\label{integralI}
I(x) \equiv \int_{0}^{2\pi} \frac{\dd k }{2\pi} \frac{\ex{\ii k x}}{1-\xi \cos k}.
\end{equation}
This integral is the generating function associated to the propagator of a symmetric nearest-neighbor random walk  between two points of a one-dimensional lattice separated by a distance $|x|$, whose expression is well-known \cite{Hughes1995}:
\begin{equation}
\label{Iexp}
I(x) = G_1(\xi) f_1(\xi)^{|x|} = \frac{1}{\sqrt{1-\xi^2}} \left( \frac{1-\sqrt{1-\xi^2}}{\xi}\right)^{|x|}.
\end{equation}
Using the expression of $I(x)$ from Eq. (\ref{Iexp}) into Eq. (\ref{PbI}), one gets the following expression for $ \mathcal{P}_{\mathrm{b}}(k_2;\xi)$:
\begin{equation}
\label{ }
\mathcal{P}_{\mathrm{b}}(k_2;\xi) = \frac{2}{\xi} \frac{\ex{\ii k_2 s_2} f_1(\xi)^{|s_1|}}{\frac{1}{f_1(\xi)}-\cos k_2},  \\
\end{equation}
where we used the relation $2/[\xi G_1(\xi)]+f_1(\xi) = 1/f_1(\xi)$ to simplify the result. Replacing $\mathcal{P}_{\mathrm{b}}(k_2;\xi) $ by its expression in Eq. (\ref{rel_P_Pb}), we get the following expression for $\mathcal{P}(k_1,k_2;\xi)$:
\begin{equation}
\label{ }
\mathcal{P}(k_1,k_2;\xi) = \frac{1}{1-\xi \cos k_1} \left[ \ex{\ii k_1 s_1} \ex{\ii k_2 s_2}+ (\cos k_2 - \cos k_1) \frac{\ex{\ii k_2 s_2} f_1(\xi)^{|s_1|}}{\frac{1}{f_1(\xi)}-\cos k_2}   \right]
\end{equation}
In what follows, we invert this Fourier transform with respect to $k_1$ and $k_2$. We do not rename $\mathcal{P}$ for simplicity. The inversion with respect to $k_1$ yields
\begin{eqnarray}
\label{ }
\mathcal{P}(r_1,k_2;\xi) &=& \ex{\ii k_2 s_2} \int_0^{2\pi} \dd k_1 \frac{\ex{\ii k_1 (s_1-r_1)}}{1-\xi \cos k_1}   \\
&& + \frac{\ex{\ii k_2 s_2} f_1(\xi)^{|s_1|}}{\frac{1}{f_1(\xi)}-\cos k_2}  \left[   \cos k_2 \int_0^{2 \pi} \dd k_1 \frac{\ex{-\ii k_1 r_1}}{1-\xi \cos k_1} - \int_0^{2 \pi} \dd k_1 \frac{\cos k_1 \ex{-\ii k_1 r_1}}{1-\xi \cos k_1}  \right] \\
&=& \ex{\ii k_2 s_2}I(s_1-r_1) + \frac{\ex{\ii k_2 s_2} f_1(\xi)^{|s_1|}}{ \frac{1}{f_1(\xi)}-\cos k_2}  \left[   \cos k_2 \, I(r_1) - \frac{1}{\xi} K\left(\frac{1}{\xi}, r_1 \right) \right], \label{inv1inter}
\end{eqnarray}
where we define for $|a|>1$:
\begin{equation}
K(a,x) \equiv \int_0^{2\pi} \frac{\dd k}{2\pi} \frac{ \ex{-\ii k x}\cos k}{a-\cos k}.
\end{equation}
This integral can be calculated as a particular case of $I(x)$ (Eq. (\ref{integralI})), and one gets:
\begin{equation}
K(a,x) =
\begin{cases}
   \frac{a-\sqrt{a^2-1}}{\sqrt{a^2-1}}   & \text{if $x=0$}, \\
  \frac{a (a-\sqrt{a^2-1})^{|x|}}{\sqrt{a^2-1}}    & \text{if  $x\neq0$}.
\end{cases}
\label{Kexp}
\end{equation}
Therefore, using the expression of $I(x)$ (Eq. (\ref{Iexp})) in Eq. (\ref{inv1inter}), we get
\begin{equation}
\label{ }
\mathcal{P}(r_1,k_2;\xi) = \ex{\ii k_2 s_2} G_1(\xi) f_1(\xi)^{|s_1-r_1|} +  \frac{\ex{\ii k_2 s_2} f_1(\xi)^{|s_1|}}{ \frac{1}{f_1(\xi)}-\cos k_2}  \left[   \cos k_2\,  G_1(\xi) f_1(\xi)^{|r_1|} - \frac{1}{\xi} K\left(\frac{1}{\xi}, r_1 \right) \right].
\end{equation}
The inversion of $\mathcal{P}(r_1,k_2;\xi)$ with respect to $k_2$ yields the following expression for $\widehat{P}(r_1,r_2;\xi)$:
\begin{eqnarray}
\label{ }
\widehat{P}(r_1,r_2;\xi) &=& G_1(\xi) f_1(\xi)^{|s_1-r_1|} \delta_{r_2,s_2} + G_1(\xi) f_1(\xi)^{|s_1|+|r_1|} K\left( \frac{1}{f_1(\xi)}, s_2-r_2\right) \nonumber\\
&& - \frac{1}{\xi} K\left( \frac{1}{\xi}, r_1  \right) f_1(\xi)^{|s_1|} \frac{\left( \frac{1}{f_1(\xi)} - \sqrt{\frac{1}{f_1(\xi)^2} -1} \right)^{|s_2-r_2|}}{ \sqrt{\frac{1}{f_1(\xi)^2} -1} }, \label{P2combEM1}
\end{eqnarray}
where we used again the result from Eq. (\ref{integralI}) and (\ref{Iexp}) to calculate the following integral: 
\begin{equation}
\label{Jexp}
J(a,x)\equiv \int_0^{2\pi}  \frac{\dd k}{2\pi}  \frac{\ex{-\ii k x}}{a-\cos k} = \frac{(a-\sqrt{a^2-1})^{|x|}}{\sqrt{a^2-1}}.
\end{equation}
With a simple calculation and using the expressions of $f_1(\xi)$ (Eq. (\ref{f1})), $G_2(\xi)$ (Eq. (\ref{G2})) and $f_2(\xi)$ (Eq. (\ref{f2})), it is easy to show that the following relations hold:
\begin{eqnarray}
\sqrt{\frac{1}{f_1(\xi)^2} -1} & = & \frac{2}{\xi G_2(\xi)} \label{inter1}\\
 \frac{1}{f_1(\xi)} - \sqrt{\frac{1}{f_1(\xi)^2} -1} & = & f_2(\xi). \label{inter2}
\end{eqnarray}
The expression of $K$ (Eq. (\ref{Kexp})) also yields:
\begin{equation}
\label{Kcalc1}
\frac{1}{\xi} K\left(  \frac{1}{\xi},r_1  \right)=
\begin{cases}
  G_1(\xi) f_1(\xi)    & \text{if $r_1=0$}, \\
\frac{1}{\xi} G_1(\xi) f_1(\xi)^{|r_1|}      & \text{if $r_1\neq0$},
\end{cases}
\end{equation}
and
\begin{equation}
\label{Kcalc2}
\frac{1}{\xi} K\left(  \frac{1}{f_1(\xi)},s_2-r_2  \right)=
\begin{cases}
  \frac{1}{2} \xi  G_2(\xi)f_2(\xi)  & \text{if $r_2=s_2$}, \\
  \frac{1}{2f_1(\xi)} \xi G_2(\xi) f_2(\xi)^{|s_2-r_2|}      & \text{if $r_2\neq s_2$}.
\end{cases}
\end{equation}
Finally, using Eqs. (\ref{inter1}), (\ref{inter2}), (\ref{Kcalc1}) and (\ref{Kcalc2}) in order to simplify Eq. (\ref{P2combEM1}), and with simple algebra, we obtain the following expression for the propagator $\widehat{P}(r_1,r_2;\xi)$:
\begin{equation}
\label{prop2comb}
\widehat{P}(r_1,r_2;\xi) = \delta_{r_2,s_2} G_1(\xi) f_1(\xi)^{|s_1-r_1|} + \frac{1}{2} G_1(\xi) G_2(\xi) f_1(\xi)^{|s_1|+|r_1|} f_2(\xi)^{|s_2-r_2|}\left[ \frac{1}{G_1(\xi)} (1+\delta_{r_1,0}) - \frac{2}{G_2(\xi)} \delta_{r_2,s_2}   \right],
\end{equation}
where $G_1(\xi)$ was defined in Section \ref{endingpointbb}, $G_2(\xi)$ is given by Eq. (\ref{G2}),  $f_1(\xi)$ by Eq. (\ref{f1}) and  $f_2(\xi)$ by Eq. (\ref{f2}).  This expression holds for \emph{any} starting point and arrival points. Therefore, we found a single expression that accounts for the different cases we considered by the renewal approach (see Eqs. (\ref{P_arrive_bb}), (\ref{reverse}), (\ref{arbitraryC2}) and (\ref{arbitraryC22})). The master equation approach then provides a single and compact expression for the propagator associated to a random walk on $\mathbf{C}_2$ with arbitrary starting and arrival points. 

In the next section, we show that this method can be extended to study the propagators associated to a random walk on a two-dimensional comb with teeth of finite length.

\subsection{Two-dimensional comb with finite teeth}

In this section, we consider a two-dimensional comb and assume that its teeth are finite, so that $r_1 \in \llbracket 0,L-1 \rrbracket$. We assume periodic boundary conditions, such that for any value of $r_2$ the sites $(0,r_2)$ and $(L,r_2)$ coincide.

The evolution of the random walker is given by the following master equations:
 \begin{itemize}
  \item for $r_1=1$ and $r_1=L-1$:
  \begin{equation}
    \label{EMPBC1}
P_{t+1}(r_1,r_2)=\frac{1}{2}[P_t(r_1+1,r_2)+P_t(r_1-1,r_2)],
\end{equation}
  \item for $r_1=1$:
  \begin{equation}
    \label{EMPBC2}
  P_{t+1}(1,r_2) =\frac{1}{4}P_t(0, r_2) +\frac{1}{2}P_t(2,r_2),
\end{equation}
  \item for $r_1=L-1$:
    \begin{equation}
      \label{EMPBC3}
  P_{t+1}(L-1,r_2) =\frac{1}{4}P_t(0, r_2) +\frac{1}{2}P_t(L-2,r_2),
\end{equation}
\item for $r_1=0$:
  \begin{equation}
  \label{EMPBClast}
 P_{t+1}(0,r_2)=\frac{1}{2}[P_t(1,r_2)+P_t(L-1,r_2)]  +\frac{1}{4}[P_t(0,r_2+1)+P_t(0,r_2-1)],
\end{equation}
\end{itemize}
with the initial condition $P_0(\rr)=\delta_{\rr,\sss}$. We define the following Fourier-Laplace transforms:
\begin{eqnarray}
\mathcal{P}_{\mathrm{b}}(k_2;\xi) & \equiv & \sum_{t=0}^\infty  \sum_{r_2=-\infty}^\infty \ex{\ii k_2 r_2} P_t(0,r_2)\xi^t, \label{def_Pb}\\
\mathcal{P}(k_1,k_2;\xi) & \equiv &  \sum_{t=0}^\infty  \sum_{r_2=-\infty}^\infty \sum_{r_1=0}^{L-1} \ex{\frac{2\ii \pi k_1 r_1}{L}} \ex{\ii k_2 r_2} P_t(r_1,r_2)\xi^t .
\end{eqnarray}
Multipliying Eq. (\ref{EMPBClast}) by $\xi^t  \ex{\frac{2\ii \pi k_1 r_1}{L}} \ex{\ii k_2 r_2} $, summing over $r_2$ and $r_1$ and using Eqs. (\ref{EMPBC1})-(\ref{EMPBC3}), we obtain the following relation between $\mathcal{P}$ and $\mathcal{P}_{\mathrm{b}}$:
\begin{equation}
\label{rel_P_Pb_PBC}
\mathcal{P}(k_1,k_2;\xi)=\frac{\ex{\frac{2\ii \pi k_1 s_1}{L}}\ex{\ii k_2 s_2}+\frac{\xi}{2}\left(\cos k_2 -\cos\frac{2\pi k_1}{L}\right)\mathcal{P}_{\mathrm{b}}(k_2;\xi)}{1-\xi \cos\frac{2\pi k_1}{L}}.
\end{equation}
Noticing that $\mathcal{P}$ and $\mathcal{P}_{\mathrm{b}}$ are related through:
\begin{equation}
\label{ }
\mathcal{P}_{\mathrm{b}}(k_2;\xi)=\frac{1}{L} \sum_{k_1=0}^{L-1} \mathcal{P}(k_1,k_2;\xi),
\end{equation}
we can sum Eq. (\ref{rel_P_Pb_PBC}) over $k_1$ to obtain the self-consistent equation satisfied by $\mathcal{P}_{\mathrm{b}}(k_2;\xi)$:
\begin{equation}
\label{Pbselfcons}
\mathcal{P}_{\mathrm{b}}(k_2;\xi) = \ex{\ii k_2 s_2} S(s_1)+\frac{\xi}{2} \cos k_2 \mathcal{P}_{\mathrm{b}}(k_2;\xi) S(0) - \frac{\xi}{2} \mathcal{P}_{\mathrm{b}}(k_2;\xi) S(1),
\end{equation}
where we defined
\begin{equation}
\label{ }
S(s_1) \equiv \frac{1}{L} \sum_{k_1=0}^{L-1} \frac{\ex{\frac{2\ii\pi k_1s_1 }{L}}}{1- \xi \cos \frac{2 \pi k_1}{L}}.
\end{equation}
In Appendix \ref{calculST}, we show that the sums $S(s_1)$ can be rewritten as follows:
\begin{equation}
\label{ }
S(s_1)=\frac{G_1(\xi)}{1-f_1(\xi)^L} \left[  f_1(\xi)^{|s_1|} + f_1(\xi)^{L-|s_1|}  \right],
\end{equation}
where we use again the quantities $G_1(\xi)$ and $f_1(\xi)$ that were defined in Section \ref{endingpointbb}. From Eq. (\ref{Pbselfcons}) we obtain the expression for $\mathcal{P}_{\mathrm{b}}(k_2;\xi)$ that we use in  Eq. (\ref{rel_P_Pb_PBC}) to obtain the following expression for $\mathcal{P}(k_1,k_2;\xi)$:
\begin{equation}
\label{ }
\mathcal{P}(k_1,k_2;\xi)=\frac{\ex{\frac{2\ii\pi k_1 s_1}{L}} \ex{\ii k_2 s_2}}{1-\xi \cos \frac{2\pi k_1}{L}}+\frac{\frac{\xi}{2} \left( \cos k_2 -\cos \frac{2 \pi k_1}{L}\right)}{1-\xi \cos \frac{2\pi k_1}{L}} \frac{\ex{\ii k_2 s_2} S(s_1)}{1+ \frac{\xi}{2} [S(1)-\cos k_2 S(0)]}.
\end{equation}
We can finally compute the inverse Fourier transform with respect to $k_1$ and $k_2$ and we finally obtain the Laplace transform of the generic propagator:
\begin{equation}
\label{propPBC}
\widehat{P}(r_1,r_2;\xi)= \delta_{s_2,r_2} S(s_1-r_1) + \frac{S(s_1)S(r_1)}{S(0)} K\left( \frac{1}{f_1'(\xi)}, s_2-r_2\right) - \frac{S(s_1)T(r_1)}{S(0)} J\left( \frac{1}{f_1'(\xi)}, s_2-r_2\right),
\end{equation}
where we defined $f_1'(\xi)$ by
\begin{equation}
\label{ }
\frac{1}{f_1'(\xi)} \equiv \frac{1}{1+f_1(\xi)^L} \left[  \frac{2}{\xi G_1(\xi)} (1-f_1(\xi)^L) +f_1(\xi)+f_1(\xi)^{L-1}\right],
\end{equation}
and the sums $T(r_1)$ by
\begin{equation}
\label{ }
T(r_1)=\frac{1}{L} \sum_{k_1=0}^{L-1} \frac{\ex{\frac{-2\ii \pi k_1 r_1}{L}}\cos \frac{2 \pi k_1}{L}}{1-\xi \cos \frac{2\pi k_1}{L}}.
\end{equation}
We present in Appendix \ref{calculST} an explicit calculation of the sums $T(r_1)$, which results in the following expression:
\begin{equation}
\label{ }
T(r_1)=
\begin{cases}
   \frac{G_1(\xi)}{2[1-f_1(\xi)^L]} \left[ f_1(\xi)^{|r_1| +1} + f_1(\xi)^{|r_1| -1}  +f_1(\xi)^{L-|r_1| -1} +f_1(\xi)^{L-|r_1| +1}  \right]   & \text{if $r_1\neq0$}, \\
   \frac{G_1(\xi)}{1-f_1(\xi)^L} \left[ f_1(\xi) +f_1(\xi)^{L -1} \right]     & \text{if $r_1=0$}.
\end{cases}
\end{equation}
Consequently, recalling the expressions of the integrals $K$ (Eq. (\ref{Kexp})) and $J$ (Eq. (\ref{Jexp})), Eq. (\ref{propPBC}) provides an explicit expression of the Laplace transform of the propagator associated to a random walk between to arbitrary points of a  two-dimensional comb with finite teeth and periodic boundary conditions. It is straightforward to check that one retrieves the result from the previous section (Eq. (\ref{prop2comb})) by taking the limit where $L\to\infty$ in Eq. (\ref{propPBC}).

\subsection{Three-dimensional comb}

We now apply the master equation approach to the case of a three-dimensional comb (see Fig. \ref{fig:combs}). On this structure, three cases can arise depending on the position of the random walker:
\begin{enumerate}
  \item if $r_1\neq0$, each neighboring site is chosen with the same probability $1/2$,
  \item it $r_1=0$ and $r_2\neq0$, each neighboring site is chosen with the same probability $1/4$,
  \item  it $r_1=0$ and $r_2=0$, each neighboring site is chosen with the same probability $1/6$.
\end{enumerate}
The evolution of the propagator $P_t(\rr)$ (where the dependence over the starting site $\sss$ is not explicitly specified) is then given by the following master equations together with the initial condition $P_0(\rr)=\delta_{\rr,\sss}$:
\begin{itemize}
  \item for $|r_1|>1$:
  \begin{equation}
\label{EMC3first}
P_{t+1}(\rr)=\frac{1}{2} [P_t(\rr+\ee_1)+P_t(\rr-\ee_1)]
\end{equation}
  \item for $|r_1|=1$ and $|r_2|>0$:
  \begin{equation}
\label{ }
P_{t+1}(\rr)=\frac{1}{4} P_t(\rr-r_1\ee_1)+\frac{1}{2}P_t(\rr+r_1\ee_1)
\end{equation}
  \item for $|r_1|=1$ and $r_2=0$:
  \begin{equation}
\label{ }
P_{t+1}(\rr)=\frac{1}{6} P_t(\rr-r_1\ee_1)+\frac{1}{2}P_t(\rr+r_1\ee_1)
\end{equation}
  \item for $r_1=0$ and $|r_2|>1$:
  \begin{equation}
\label{ }
P_{t+1}(\rr)=\frac{1}{2} [P_t(\rr+\ee_1)+P_t(\rr-\ee_1)]+\frac{1}{4} [P_t(\rr+\ee_2)+P_t(\rr-\ee_2)]
\end{equation}
  \item for $r_1=0$ and $|r_2|=1$:
  \begin{equation}
\label{ }
P_{t+1}(\rr)=\frac{1}{2} [P_t(\rr+\ee_1)+P_t(\rr-\ee_1)]+\frac{1}{4} P_t(\rr+r_2\ee_2)+\frac{1}{6}P_t(\rr-r_2\ee_2)
\end{equation}
  \item for $r_1=0$ and $|r_2|=0$:
  \begin{equation}
\label{EMC3last}
P_{t+1}(\rr)=\frac{1}{2} [P_t(\rr+\ee_1)+P_t(\rr-\ee_1)]+\frac{1}{4}[ P_t(\rr+\ee_2)+P_t(\rr-\ee_2)]+\frac{1}{6}[ P_t(\rr+\ee_3)+P_t(\rr-\ee_3)]
\end{equation}
\end{itemize}
Note that, more generally, a master equation description of a random walk on a $d$-dimensional comb would yield a set of $d(d+1)/2$ distinct equations. 

We define the following Fourier-Laplace transforms:
\begin{eqnarray}
\mathcal{P}(\kk;\xi) & = & \sum_{t=0}^\infty \sum_{\rr} \xi^t \ex{\ii \kk\cdot\rr} P_t(\rr), \label{defPC3}\\
\mathcal{P}_2(k_2,k_3;\xi) & = & \sum_{t=0}^\infty \sum_{r_2,r_3=-\infty}^\infty \xi^t \ex{\ii(k_2 r_2+ k_3 r_3)} P_t(0,r_2,r_3), \\
\mathcal{P}_3(k_3;\xi) & = & \sum_{t=0}^\infty \sum_{r_3=-\infty}^\infty \xi^t \ex{\ii k_3 r_3} P_t(0,0,r_3).
\end{eqnarray}
From the definition of $\mathcal{P}$ (Eq. (\ref{defPC3})) and the master equations (Eqs. (\ref{EMC3first}) to (\ref{EMC3last})), we obtain the following relation between $\mathcal{P}$, $\mathcal{P}_2$ and $\mathcal{P}_3$:
\begin{equation}
\label{PC3}
\mathcal{P}(\kk;\xi) = \frac{1}{1-\xi\cos k_1} \left[ \ex{\ii \kk\cdot \sss} + \frac{\xi}{2} (\cos k_2 - \cos k_1) \mathcal{P}_2(k_2,k_3;\xi) + \frac{\xi}{3} \left( \cos k_3 - \frac{1}{2}\cos k_1 \right)   \mathcal{P}_3(k_3;\xi)\right].
\end{equation}
Integrating this relation over the variable $k_1$ and using the relation $\int_0^{2\pi} \dd k_1 \mathcal{P}(\kk;\xi)/(2\pi) = \mathcal{P}_2(k_2,k_3;\xi)$, we obtain the following relation between $\mathcal{P}_2$ and $\mathcal{P}_3$:
\begin{equation}
\label{P2C3}
\mathcal{P}_2(k_2,k_3;\xi)= \frac{2}{\xi} \frac{1}{\frac{G_1(\xi)}{f_1(\xi)}-\cos k_2} \left\{  \ex{\ii(k_2 s_2+k_3 s_3)} G_1(\xi) f_1(\xi)^{|s_1|} + \frac{\xi}{3} \mathcal{P}_3(k_3;\xi) \left[\cos k_3 - \frac{1}{2}G_1(\xi)f_1(\xi) \right]\right\}.
\end{equation}
Additionally, integrating this relation over the variable $k_2$ and using $\int_0^{2\pi} \dd k_2 \mathcal{P}_2(k_2,k_3;\xi)/(2\pi) = \mathcal{P}_3(k_3;\xi)$, we obtain the following closed expression for $\mathcal{P}_3(k_3;\xi)$:
\begin{equation}
\label{P3C3exp}
\mathcal{P}_3(k_3;\xi)=\frac{\frac{2}{\xi} \ex{\ii k_3 s_3} G_1(\xi) f_1(\xi)^{|s_1|}J(G_1(\xi)/f_1(\xi),s_1)}{1-\frac{2}{3}\left[ \cos k_3 - \frac{1}{2}G_1(\xi) f_1(\xi)  \right]  J(G_1(\xi)/f_1(\xi),0) }.
\end{equation}
Replacing $\mathcal{P}_3(k_3;\xi)$ by this expression in Eq. (\ref{P2C3}), we obtain the following expression for $\mathcal{P}_2(k_2,k_3;\xi)$:
\begin{eqnarray}
\label{P2C3exp}
\mathcal{P}_2(k_2,k_3;\xi)&=&\frac{2}{\xi}\frac{1}{\frac{G_1(\xi)}{f_1(\xi)}-\cos k_2} \left\{ \ex{\ii k_2 s_2} \ex{\ii k_3 s_3} G_1(\xi) f_1(\xi)^{|s_1|} \right.\nonumber\\
&&\left.+\frac{\frac{2}{\xi} \ex{\ii k_3 s_3} G_1(\xi) f_1(\xi)^{|s_1|}J(G_1(\xi)/f_1(\xi),s_1)\left[ \cos k_3 - \frac{1}{2} G_1(\xi) f_1(\xi)  \right]}{1-\frac{2}{3}\left[ \cos k_3 - \frac{1}{2}G_1(\xi) f_1(\xi)  \right]  J(G_1(\xi)/f_1(\xi),0) } \right\}
\end{eqnarray}
Finally, the relation between $\mathcal{P}$, $\mathcal{P}_2$ and $\mathcal{P}_3$ (Eq. (\ref{PC3})), together with the expressions for $\mathcal{P}_2$ (Eq. (\ref{P2C3exp})) and $\mathcal{P}_3$ (Eq. (\ref{P3C3exp})) yield an explicit expression for the Fourier-Laplace transform $\mathcal{P}(\kk;\xi)$, that can be inverted to obtain a single expression for the propagator $P_t(\rr)$ valid for any starting and arrival points.

Once again, the master equation description appears to be a powerful method as it allows to compute the propagators of a random walk on a three-dimensional comb.

\section{Conclusion}

In this paper, we studied diffusion on comb lattices of arbitrary dimension. More precisely, relying on the treelike structure of these lattices and making use of renewal equations and first-passage properties, we computed the Laplace transforms of the propagators (namely the probability for a random walker to be at a given site at a given time knowing its starting point) in both cases where the shortest path from the initial to the final point contains at least one point from the backbone or not. We obtained explicit and closed formulae, valid for comb lattices of arbitrary dimension.

We then proposed an alternative derivation of these quantities relying on a master equation of the problem. We obtained the Laplace transforms of the propagators which are given by a single and simple formula for the case of two-dimensional combs. This method was then extended to study the case where the teeth of the two-dimensional comb are finite, and to obtain the Fourier-Laplace transforms of the propagators in the case of a three-dimensional comb.

The first method allowed us to consider a specific class of random walks on comb lattices of arbitrary dimension, and obtained generic expressions for the propagators and first-passage time densities associated to these random walks. The second method  is very efficient to study random walks with arbitrary starting and ending points as long as the dimension of the comb lattice is not too large. The latter could be used to study more complex random walks on comb-like structures, that would involve drift, defective sites or finite teeth with reflexive boundary conditions. The study of such random walks could give a new insight into transport phenomena encountered  in complex and disordered systems  of physical or biological inspiration.

\section*{Acknowledgments}

PI acknowledges financial support from the U.S. National Science Foundation under MRSEC Grant No. DMR-1420620. OB acknowledges financial support from the European Research Council Starting Grant No. FPTOpt-277998.

\appendix

\section{Calculation of the sums  $S(x)$ and $T(x)$}
\label{calculST}

In this appendix, we compute the sum $S(x)$ defined by
\begin{equation}
\label{ }
S(x) = \frac{1}{L} \sum_{k=0}^{L-1} \frac{\ex{\frac{2\ii\pi k x }{L}}}{1- \xi \cos \frac{2 \pi k}{L}}.
\end{equation}
Note that $S(x)$ is symmetric with respect to $x$. We consider the case $x\geq0$ with no loss of generality. We first rewrite the sum as
\begin{equation}
\label{ }
S(x) = \frac{1}{L} \sum_{k=0}^{L-1} \frac{\left(\ex{\frac{2\ii\pi k }{L}}\right)^x}{1- \frac{\xi}{2} \left( \ex{\frac{2\ii\pi k }{L}}+\ex{\frac{-2\ii\pi k }{L}}\right)},
\end{equation}
and convert it into a sum over the roots of unity $\zeta$ satisfying $\zeta^L=1$:
\begin{equation}
\label{ }
S(x) = \frac{1}{L} \sum_{\zeta^L=1}  \frac{\zeta^x}{1- \frac{\xi}{2} \left( \zeta + \frac{1}{\zeta}\right)} = -\frac{2}{L\xi} \sum_{\zeta^L=1} \frac{\zeta^{x+1}}{\zeta^2-\frac{2}{\xi} \zeta+1}.
\end{equation}
The denominator can be factorized as follows:
\begin{equation}
\label{ }
\zeta^2-\frac{2}{\xi} \zeta+1 = \left( \zeta - \frac{1}{f_1(\xi)} \right) (\zeta-f_1(\xi))
\end{equation}
where $f_1(\xi)$ was defined in Eq. (\ref{f1}) and is the Laplace transform of a FPTD related to a random walk on an infinite one-dimensional lattice. With a partial-fraction decomposition, we rewrite $S(x)$ as
\begin{equation}
\label{ }
S(x)=- \frac{2}{L \xi} \frac{1}{f_1(\xi)-\frac{1}{f_1(\xi)}} \sum_{\zeta^L=1}\left[ \frac{\zeta^{x+1}}{\zeta-f_1(\xi)} - \frac{\zeta^{x+1}}{\zeta-\frac{1}{f_1(\xi)}}  \right].
\end{equation}
The sums over the roots of unity are then evaluated using the following relation \cite{Gessel1997}, which holds for $n$ integer, $r\in \llbracket 1,n \rrbracket$ and $a\neq1$:
\begin{equation}
\label{Gessel}
\sum_{\zeta^n=1} \frac{\zeta^r}{\zeta-a} = 
\begin{cases}
 n \frac{a^{r-1}}{1-a^n}     & \text{if $r\neq0$}, \\
     n \frac{a^{n-1}}{1-a^n}  & \text{if $r=0$}.
\end{cases}
.
\end{equation}
We finally obtain
\begin{equation}
\label{ }
S(x) = \frac{G_1(\xi)}{1-f_1(\xi)^L} \left[ f_1(\xi)^{|x|} +f_1(\xi)^{L-|x|}   \right],
\end{equation}
where we used the relation  $1/f_1(\xi)-f_1(\xi) =2/[\xi G_1(\xi)]$.

The sums $T(x)$ are evaluated in a similar way. Recalling their definition:
\begin{equation}
\label{ }
T(x)=\frac{1}{L} \sum_{k=0}^{L-1} \frac{\ex{\frac{-2\ii \pi k x}{L}}\cos \frac{2 \pi k}{L}}{1-\xi \cos \frac{2\pi k}{L}},
\end{equation}
we see that $T$ is symmetric with respect to $x$ and will consider the case $x\leq 0$ for convenience. Writing the sum over $k$ as a sum over the root of unity, we get
\begin{eqnarray}
\label{ }
T(x)&=&\frac{1}{L} \sum_{\zeta^L=1}   \frac{\frac{1}{2}\zeta^{-x}\left( \zeta+\frac{1}{\zeta}\right)}{1-\frac{\xi}{2} \left(  \zeta + \frac{1}{\zeta} \right)}\\
&=& -\frac{1}{\xi L}\frac{1}{f_1(\xi)-\frac{1}{f_1(\xi)}} \sum_{\zeta^L=1} (\zeta^{-x+2}+\zeta^{-x})\left[ \frac{1}{\zeta-f_1(\xi)} - \frac{1}{\zeta-\frac{1}{f_1(\xi)}}  \right].
\end{eqnarray}
Finally, using again Eq. (\ref{Gessel}) and generalizing for any sign of $x$, we get
\begin{equation}
\label{ }
T(x)=
\begin{cases}
   \frac{G_1(\xi)}{2[1-f_1(\xi)^L]} \left[ f_1(\xi)^{|x| +1} + f_1(\xi)^{|x| -1}  +f_1(\xi)^{L-|x| -1} +f_1(\xi)^{L-|x| +1}  \right]   & \text{if $x\neq0$}, \\
   \frac{G_1(\xi)}{1-f_1(\xi)^L} \left[ f_1(\xi) +f_1(\xi)^{L -1} \right]     & \text{if $x=0$}.
\end{cases}
\end{equation}


\end{document}